\documentstyle[aps,prl,epsf]{revtex}
\begin{document}
\advance\textheight by 0.5in
\advance\topmargin by -0.2in
\draft

\twocolumn[\hsize\textwidth\columnwidth\hsize\csname@twocolumnfalse%
\endcsname

\title{Quasi-static crack propagation in heterogeneous media}
\author{Sharad Ramanathan, Deniz Erta\c s and Daniel S. Fisher} 
\address{Lyman Laboratory of Physics, Harvard University, Cambridge, 
Massachusetts 02138}

\date{November 20, 1996} 

\maketitle

\begin{abstract}
The dynamics of a single crack moving through  a heterogeneous medium
is studied in the quasi-static approximation. Equations of motion for
the crack front are formulated and the resulting scaling behaviour analyzed.
In a model scalar system and for mode III (tearing) cracks, the crack surface 
is found to be self affine with a roughness exponent of $\zeta=1/2$. But in 
the usual experimental case of mode I (tensile) cracks, local mode preference 
causes the crack surface to be only logarithmically rough, quite unlike those 
seen in experiments. The effects of residual stresses are considered and 
found, potentially, to lead to increased crack surface roughness.
But it appears likely that elastic wave propagation effects may be needed 
to explain the very rough crack surfaces observed experimentally.
\end{abstract}
\vskip -0.6 truein
]

Since the work of Mandelbrot, Passoja and 
Paullay\cite{Mandel},
many experimental\cite{Bou2,ceramics,Rock} and numerical\cite{MD} 
studies on geometrical properties of
crack surfaces and crack fronts in heterogeneous materials have 
demonstrated their self-affine scaling properties. 
Cracks, in a large variety of materials from aluminum alloys\cite{Bou2} 
to ceramics\cite{ceramics} and rock\cite{Rock}, seem to leave 
behind surfaces whose height variations are characterized by a 
roughness exponent $\zeta\approx 0.8$. 
In the meantime, there have been significant advances in the understanding 
of the dynamics of interfaces and lines pinned by quenched random 
impurities\cite{NSTL,NF,CL}. Potentially, these might shed light on the 
problem of crack surfaces and fronts\cite{Bou1}, since a crack surface 
can be viewed as the trace left behind by the crack front 
as it traverses the sample. But such theoretical analyses of crack surfaces 
have so far neglected the crucial long range effects of elasticity. 
In particular, the deformations of a crack front change stresses 
at distant points on the front\cite{Rice1985}. Moreover, the  
stresses are also modified by the boundary conditions on the 
rough crack surface, which give rise to dependence on the history 
of the crack front. As the crack moves, it relieves both the elastic stresses
generated by the  external loading and those due to the presence
of quenched impurities such as inclusions, micro-cracks or
dislocations. Because of elasticity, these defects give
rise to long-range correlated randomness in the equations of motion. 

All of these effects influence the macroscopic scaling behavior of the 
crack surface and have to be included in a realistic model of crack 
propagation. Unfortunately, the full elastodynamic problem 
is very difficult. Indeed, even the criteria that determine the local 
direction of advance of crack front are not well understood\cite{crackdir}.

In this paper, we present the results of a study of a crack 
propagating 
through a heterogeneous medium in the quasi-static approximation. 
We first consider a simplified {\it scalar} elasticity theory and then the 
real case of vectorial elasticity. We use the symmetries of the system 
and elasticity theory to determine phenomenological equations of motion. 

Our conclusions are, unfortunately, that for the principal situation
of experimental interest of tensile (mode I) cracks 
[and also shear (mode II) cracks] the predicted roughness of the crack 
front is only logarithmic rather than a power law of the length scale. 
Tearing (mode III) cracks are substantially rougher; however, these 
tend not to be stable experimentally and thus the applicability of the 
results is questionable. In all cases, random long wavelength residual 
stresses can increase the roughness; the effects of these are discussed 
briefly.

\begin{figure}
\narrowtext \epsfxsize=2.9truein \vbox{\hskip 0.15truein
\epsffile{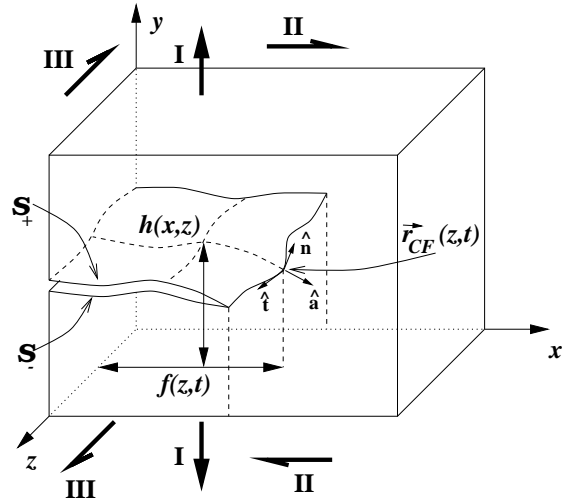}}
\medskip
\caption{Crack propagating through a heterogeneous medium. The
$x$ co-ordinate of the front is $f(z,t)$ and the free crack
surfaces ${\cal S}_+,{\cal S}_-$ are located at $y=h(x,z)$. 
The local tangent, forward  and
normal vectors ${\bf\hat t}, {\bf\hat a}, {\bf\hat n}$
at a point ${\bf r}_{CF}$ on the front are also shown as are the directions
of mode I (tensile), mode II (shear) and mode III (tearing) loading.}
\label{crack3d}
\end{figure}

The geometry of the crack is shown in Fig.~\ref{crack3d}. 
The crack front is oriented along the $z-$axis and moves in the 
positive 
$x$ direction. The $x$ (in-plane) coordinate of the front 
at time $t$, $x_{c}(z,t)=f(z,t)$  is assumed to be single valued 
in $z$. As the crack moves, it leaves behind  free surfaces ${\cal S}_+$
and ${\cal S}_-$ on either side of the crack, located in body coordinates 
at $y=h(x,z)$ for $x<f(z,t)$. Thus, the instantaneous position of the 
front is given by the curve ${\bf r}_{CF}(z,t)\equiv f(z,t){\bf \hat{x}}
+h[f(z,t),z]{\bf \hat{y}}+z{\bf \hat {z}}$.
We assume that the crack motion is quasi-static, so that at each instant,
the scalar displacement field $u(x,y,z)$ satisfies Laplace's equation with 
the boundary condition that the normal stresses on the  surfaces  
${\cal S}_+$ and ${\cal S}_-$ vanish. The sample is loaded far away from 
the crack, such that 
\begin{equation}
u {\approx} \frac{K^{\infty}y}{2M\sqrt{\pi}}
\sqrt\frac{\sqrt{x^{2}+y^{2}}-x}{y^2},
\label{bc2}
\end{equation}
for $\sqrt{x^{2}+y^{2}}$ large, with  $K^{\infty}$ the applied stress 
intensity factor and $M$ the elastic modulus. Displacement and stress 
fields in the medium are determined by the position of the front and 
the shape of the crack surface and these in turn influence the subsequent 
motion of the crack. We now derive general equations of motion for the time 
evolution of the crack and analyze them to determine scaling exponents that 
characterize the roughness of crack fronts and surfaces.  

A local coordinate system can be defined at a point ${\bf r}_{CF}$
on the  crack front, with axes along the tangent 
${\bf \hat t}\equiv \partial_{z}{\bf r}_{CF}/|\partial_{z}{\bf r}_{CF}|$, 
 forward direction ${\bf \hat a}\equiv \partial^{2}_{z}{\bf r}_{CF}/
|\partial^{2}_{z}{\bf r}_{CF}|$,
and normal ${\bf \hat n}\equiv{\bf \hat t}\times {\bf \hat a}$ 
(see Fig.~\ref{crack3d}).
The direction in which the crack front moves should be
governed by the angular dependence of the
stress and the local yield stress near the crack tip. The
stress field close to the crack tip has the form
\begin{equation}
\mbox{\boldmath$\sigma$}
=\frac{K({\bf r}_{CF})}{\sqrt{2{\pi}\rho}}
[\cos(\theta/2){\bf\hat{n}}-\sin(\theta/2){{\bf\hat{a}}}] 
+ S_{a}({\bf r}_{CF}){\bf\hat{a}}+{\cal O}(\sqrt{\rho}),
\label{stress}
\end{equation}
with $\rho$ the distance to the crack front and $\theta$ the angle with 
respect to ${\bf\hat{a}}$ in the $({\bf\hat{a}}$-${\bf\hat{n}})$ plane. 
The local 
stress intensity factor is $K({\bf r}_{CF})$ and ${\bf{S}}$ is the finite 
part of the stress whose normal component $S_{n}$ must vanish. 
The divergent part of the stress field does not break the reflection 
symmetry with respect to the crack front plane (spanned by $\hat{\bf t}$ 
and  $\hat{\bf a}$), so the crack should move in the forward direction 
$\hat{\bf a}$\ leaving behind a crack surface that is smooth on the 
smallest length scales.

As the crack front advances, an elastic energy 
$
G(K^{\infty},{\bf r}_{CF},\{f\},\{h\})=
\frac{K^{2}({\bf{r}}_{CF})}{M}
$
per unit area of 
generated crack surface is released by the medium.
In order to create the new crack surface, a surface energy 
$\Gamma({\bf r}_{CF})$ per unit area is required. This is 
a local material property. We will assume that the  zone with non-linear 
deformations, plasticity or microscopic side branches is small 
and that linear elasticity theory is applicable 
outside of this region; $\Gamma$ will include  the energy 
to break bonds, as well as that dissipated as a result of plastic 
flow etc. Under quasi-static conditions,
the forward velocity $v_{a}\equiv{\bf \hat a}\cdot \partial_{t}{\bf r}_{CF}$ 
of the crack front is proportional to
the {\it excess} elastic energy released, i. e.,
\begin{equation}
\mu^{-1} v_{a}({\bf{r}}_{CF})=G(K^{\infty},{\bf r}_{CF},\{h\},\{f\})
-\Gamma({\bf r}_{CF}),
\label{gme}
\end{equation}
where $\mu$ is the effective mobility of the crack front. If $\mu$ is much smaller than $(c\Gamma)^{-1}$, with $c$ the speed of sound, the quasistatic approximation should be good except on very long length scales.
 
Energy considerations alone cannot determine how the direction of motion of 
the  crack front changes. Instead, we expect this to be determined by the 
small scale processes that break the reflection symmetry about the local 
crack surface. In particular, the finite part of the stress field near the 
crack front, $S_{a}({\bf{r}}_{CF})$ in Eq.(\ref{stress}) is antisymmetric 
about the crack surface. The competition of this with the local random 
variations in the material yield stress will determine the 
{\it curvature} $\kappa\equiv{\bf\hat{n}}\cdot\partial_a{\bf\hat{a}}$ 
of the path of the crack front as it advances:  
$S_{a}$ will tend to make the crack curve in the direction in which the 
``hoop'' stress $\sigma_{\theta}$ or the total stress $|\bf{\sigma}|$ is 
larger. For small changes in the direction of advance, this will be 
determined by the stresses near $\theta=0$; from the small $\theta$ 
behaviour of Eq.(~\ref{stress}), we thus see that $\kappa$ should be 
proportional to $-S_{a}$. But the yield stress ( or other appropriate 
fracture stress) of the material, ${\sigma_{\rm Y}}$, can vary with 
position and the crack will tend to bend in a direction in which 
${\sigma_{\rm Y}}$ is lower. We postulate that the competition between 
these effects determines the local curvature:
\begin{equation}
\kappa=-(S_{a}+b{\partial}_{n}{\sigma_{\rm Y}})/\lambda,
\label{outofplane}
\end{equation}
with $b{\sim}K_{c}^2/{{\sigma_{\rm Y}}}^{2}M^{2}$ of order the distance from 
the crack tip at which the stress is ${\sigma_{\rm Y}}$ and 
${\lambda}{\sim}K_{c}\sqrt{b}$, with $K_{c}$ the critical stress intensity 
factor for crack advance. The detailed form of Eq.(\ref{outofplane}) will 
turn out to be unimportant at long length scales.

Both $S_{a}$ and the energy release rate, $G$, can be written as a sum of 
two terms. The first terms, $G^{u}$ and ${S}^{u}_{a}$ are the energy 
release rate and the nonsingular stress near the crack front for a  
homogeneous medium with the same loading and geometry of the crack surface, 
and thus do not depend on the heterogeneities in the system. These are 
functionals of the entire shape of the front and the crack surface, as well 
as the applied load. The second terms, $G^{r}$ and ${S}^{r}_{a}$ are the 
contributions from the relief of frozen-in stresses due to the 
heterogeneities and defects in the medium as well as any elastic 
inhomogeneities.

In general, $G^{r}$ and ${S}^{r}_{a}$ will also depend on the crack shape.
 But for almost straight cracks , they will primarily depend on the
 unperturbed shape of the crack and to linear order in the randomness
this effect can be neglected. 
We can now define random variables 
\begin{eqnarray}
\chi(x,z)\equiv b{\partial}_{n}{{\sigma_{\rm Y}}}[x,h(x,z),z]
+S_{a}^{r}[x,h(x,z),z], \\
\gamma(x,z)\equiv{\Gamma}[x,h(x,z),z]-{\bar\Gamma}-G^{r}[x,h(x,z),z],
\label{gamma1}
\end{eqnarray}
on the crack surface, with ${\langle}G^{r}\rangle 
={\langle}S^{r}_{a}\rangle=0$ and ${\langle}\Gamma\rangle=\bar{\Gamma}$, 
the mean fracture toughness.
These will depend on $h(x,z)$ but to lowest order in $h$ this effect is 
not important. The means of $\gamma$ and $\chi$ are defined to be zero and their correlations are taken to be
\begin{eqnarray}
\langle\gamma(x,z)\gamma(x',z')\rangle
&=&\Delta(x-x',z-z'), \\
\langle\chi(x,z)\chi(x',z')\rangle
&=&\Upsilon(x-x',z-z').
\label{chi}
\end{eqnarray}
We first consider short range correlations where 
$\Delta$ and $\Upsilon$ $\propto \delta(x-x')\delta(z-z')$,with some short distance cut-off,
later returning to the effects of long range correlations.
The loads and the shape of the crack surface determine the stress field. 
Consequently Eqs.(\ref{gme}-\ref{chi}) determine the evolution 
of the crack front.

We have obtained a  solution for the displacement field perturbatively 
around a flat, straight crack in an isotropic, homogeneous medium that can 
be carried out order by order in $h(x,z)$ and $f(z,t)$\cite{unpublished}.
This solution combined with
Eq.(\ref{gme}) yields an equation of motion for $f(z,t)$.
To linear order, out-of-plane deviations of the crack surface do not 
contribute 
to the divergent stress just ahead of the crack tip that determines 
the energy release rate. The behavior of the in-plane displacement $f$ is 
identical
to that of a crack which is restricted to move in a plane:
\begin{eqnarray}
\mu\partial_{t}f(z,t)&=&\frac{\sqrt{G^{\infty}}}{\pi}{\cal{{P}}}\!\!
\int_{-\infty}^{\infty}dz'
\frac{f(z',t)-f(z,t)}{|z-z'|^2}\nonumber\\
& &- {\gamma}(f(z,t),z)+G^{\infty}-\bar{\Gamma},
\label{3dm}
\end{eqnarray}
 with $G^{\infty}=(K^{\infty})^{2}/M$ and ${\cal P}$ denotes the principal
 part of 
the integral.
Similarly, Eq.(\ref{outofplane}) gives us an equation for the evolution 
of $h(x,z)$. Stresses generated by in-plane fluctuations of the crack front 
do not 
break the reflection symmetry and therefore cannot affect the
curvature $\kappa$. Thus, the leading correction to
$S^{u}_a$ involves only out-of-plane fluctuations $h$.
The linearized equation for the evolution of $h$ as the front passes a 
point ${\bf r}_{CF}$
at $x=f(z,t)$ is 
\begin{eqnarray}
\lambda \partial_{x}^{2}h(x,z)& =& 
\sqrt{G^{\infty}}{\cal{P}}\!\!\int\limits_{-\infty}^{\infty}\!\!\!dx'\!\!\!
\int\limits_{-\infty}^{\infty}\!\!\!dz' 
J(x-x',z-z') h(x',z) \nonumber\\
&& -\chi(x,z),
\label{3da}
\end{eqnarray}
independent of $f$.
 The Fourier transform of the long range kernel $J$ is 
\begin{equation}
\tilde J(q,k)=(q-i\epsilon)^{3/2}Y(k/q),
\end{equation}
with $Y$ a scaling function.
 To linear order the equations of motion for $f$ and $h$ decouple. 
Therefore the crack surface is determined independently of the dynamics 
of the front, which is not true in general.

Equation (\ref{3dm}) has been studied both analytically\cite{CL} and 
numerically\cite{unpublished}. The front is arrested for small external loads,
and there is a critical load $K^{\infty}_c$ 
(corresponding to $G^{\infty}_c$)
at which the crack just begins to move.
For a load $K^{\infty}$ slightly above this threshold, the 
average velocity of the front scales as 
$v\sim(K^{\infty}-K^{\infty}_ c)^{\beta}$.
The motion of the front is rather jerky with 
fluctuations in the velocity 
correlated up to a distance $\xi$, which diverges
at threshold like $\xi\sim(K^{\infty}-K^{\infty}_ c)^{-\nu}$.
At length scales smaller than $\xi$, the in-plane front profile  $f(z,t)$
is self-affine with
$
\langle[f(z,t)-f(z',t)]^{2}\rangle{\sim}|z-z'|^{2\zeta_{f}}.
$
A renormalization group $\epsilon$-expansion and numerical simulations
suggest
$ 
\zeta_{f}=1/3,\; \nu=3/2,\; \beta\approx 7/9.
$
At length scales larger than $\xi$ (or well above threshold), 
the crack front  moves more uniformly, and its in-plane
fluctuations scale as
\begin{equation}
\langle[f(z,t)-f(z',t)]^{2}\rangle{\sim}\log|z-z'|,
\end{equation}
as can be seen by Fourier transforming Eq.(\ref{3dm}).

In order to calculate scaling properties of the crack surface,
we next analyze Eq.(\ref{3da}).
At length scales larger than $b$, the left hand
 side of Eq.(\ref{3da}) becomes negligible and the direction 
in which the crack moves is determined by the competition between the 
non singular stress and the material properties near the crack front. 
Thus, beyond this length scale our results will be the same 
if we alternatively assume that the crack tip moves in a direction in which
 the stress $\sigma_{\theta}$ exceeds the local material yield stress 
furthest from the crack front.
The crack surface roughness thus has the scaling form
\begin{equation}
\langle[h(x,z)-h(x',z')]^2\rangle \sim |z-z'|^{2\zeta_{h}}
H\left(|x-x'|/|z-z'|\right),
\end{equation} 
with $H$ a scaling function.
The crack surface is anisotropic but with the same roughness 
exponent of $\zeta_{h}=0.5$ in both the $x$ and the $z$ directions.

We now consider the real case of vectorial elasticity.
Near the crack front, the stresses have the form
\begin{equation}
\sigma_{ij}=\sqrt{\frac{1}{2{\pi\rho}}}\{K_{\rm I}\Sigma^{\rm I}_{ij}(\theta)
+K_{\rm II}\Sigma^{\rm II}_{ij}(\theta)+K_{\rm III}\Sigma^{\rm III}_{ij}
(\theta)\},
\end{equation}
where the stress intensity factors $K_{\rm I},K_{\rm II}$ and $K_{\rm III}$ 
are associated, respectively, with discontinuities across the crack surface 
of displacements in the ${\bf\hat{n}}$ (tensile loading), ${\bf\hat{a}}$ 
(shear loading) and ${\bf\hat{t}}$ (tear loading) directions; and the 
$\Sigma^{\mu}_{ij}$ are universal functions.
If the crack is loaded far away purely in mode III, then out of plane
$z$-independent deformations $h(x)$ in the crack surface will not mix in the 
other modes near the crack front. The direction in which the crack progresses will then, 
as for the scalar case, be determined by the finite non-singular parts 
of the stress near the tip. The roughness exponents should thus be the same 
as the scalar case although the scaling functions will be different. 
Unfortunately, mode III cracks tend to be unstable\cite{ortiz}, so the applicability of 
these results is questionable.

 The primary situation of experimental interest is mode I (tensile) loading. 
In this case, once the crack wanders out of plane, the local
$K_{\rm II}$ becomes non-zero, and with $z$-dependent distortions of the 
crack, so does $K_{\rm III}$. To linear order, $ K_{\rm II}$
and $K_{\rm III}$ are functionals of $h$ while $K_{\rm I}$ only depends on  
$f$.
 As the crack moves, the change in the energy release rate $G=
{\sum_{\mu}}\frac{K_{\mu}^2}{M_{\mu}}$, where $M_{\mu}$ are the appropriate 
elastic constants \cite{lawn}, is dominated by ${\delta}K_{\rm I}$.
Therefore, as in the scalar case, the equation of motion of the in plane 
displacement will be independent of the out of plane displacement. 
The roughness of the front should be similar to the scalar case, but the out 
of plane roughness is very different.
Following various authors, with the expectation that the local dynamics will 
be determined
by the dominant terms consistent with symmetry, we assume that the crack tip 
locally prefers mode I loading\cite{cotterell,Sethna}. 
This implies that the direction of motion of 
the crack tip can be found by requiring that the curvature at the crack tip, 
$\kappa$ be proportional to the angular derivative of 
the $\sigma_{\theta\theta}$ 
component of the stress tensor, the hoop stress just ahead of the crack 
tip. As in the scalar
case, this just makes sure that the crack is smooth on small length scales,
and the results on large length scales should be independent of the specific 
form of this condition.
To linear order this leads to the equation for the out of plane displacement
\begin{eqnarray}
{\lambda}\kappa&=&-\frac{K_{\rm II}({\bf{r}}_{CF})}{\sqrt{b}}
-\chi(x=f(z,t),z)\nonumber\\
&=& -\frac{1}{\sqrt{b}}[K_{\rm I}^{\infty}\partial_{x}h+K_{\rm II}^{h}]-\chi,
\label{3dvec}
\end{eqnarray}
where $K_{\rm II}^{h}$, which is linear in $h$, is the local mode II loading 
in the original $x,y,z$ coordinates, obtained from the perturbative solution 
for the displacement field. Again, $b$ is a microscopic length, 
$\lambda{\sim}K_{\rm I}^{c}\sqrt{b}$, and $\chi$ is a random variable that 
includes the effects of local variations in the yield stress and the relief 
of random residual stresses.

Once again $f$ appears only as an implicit variable and $h$ is determined 
independently of $f$. The Fourier transform of Eq.(\ref{3dvec}) involves 
a long range kernel 
\begin{equation}
\tilde{{{J}_{\rm I}}}(q,k)=|q|{{Y}_{\rm I}}(k/q).
\end{equation} 
We now see a 
crucial difference from the scalar case: $\tilde{J}_{\rm I}$ differs from 
$\tilde{J}$ 
by a factor of $1/|q|^{1/2}$,  thus the out of plane ``stiffness'' is much 
stronger in the mode I case due to the local mode selection.  The extra 
stiffness leads to only logarithmically rough crack surfaces, in striking 
contrast to what is observed in experiments. 

A study of the non-linear terms that we have neglected throughout, leads to 
the conclusion that  they will be irrelevant at long length scales for all 
the cases considered. In our quasi-static approximation it thus appears that 
the only way to get rougher surfaces is if the random properties of the 
solid have long range correlations, which we discuss next.

Residual stresses $\sigma^{r}({\bf{r}})$, which are present in the material 
before it cracks, will change the local stress intensity factors at the 
crack front for 
a given crack geometry. If $\langle\sigma^{r}({\bf{r}})\sigma^{r}({\bf{r}}')
\rangle{\sim}1/|{\bf{r}}-{\bf{r}}'|^{\alpha}$
at long distances, dimensional analysis yields additional correlations 
in the resulting $G^r$ and  $S_a^r$, of the form
$
\Delta(x,z)\sim{|z|^{1-\alpha}}\hat{\Delta}(|{x}/{z}{|}{)}$ and $
\Upsilon(x,z)\sim{|z|^{-\alpha}}\hat{\Upsilon}{(}{|}{x}/{z}{|}{)}.
$
Randomly distributed finite sized defects such as 
heterogeneities, micro-cracks or  loop dislocations
each yield stress fields falling off as $1/r^3$\cite{Ricedef} and hence 
$\alpha=3$. For the roughness of the moving front on scales larger than 
$\xi$, and the roughness of the crack surfaces in mode I , this is a 
marginal perturbation which will result in changing the logarithmic 
roughness to $(\ln)^{2}$. Near the threshold for the crack motion, such 
correlations are also marginal.
But in some disordered materials, growth processes may result in longer 
range correlations in the residual stresses. As an extreme example, one 
could consider the stress caused by segments of dislocations placed randomly 
in the solid
with the constraint that they form loops, i.e., that the dislocation density 
tensor is divergence free. This results in $\alpha=1$ which is a strongly 
relevant perturbation yielding, in the linear approximation an unphysical 
roughness exponent of $\zeta_{h}=1$ both for the front and the mode I crack 
surface.
Generally, we have $\zeta_{h}=(3-\alpha)/2$ for $\alpha<3$, thus very long range 
correlations with $\alpha{\approx}1.4$ would be needed to explain the 
experiments. 

On the basis of our results an explanation of the measured roughness in terms
of quasi-static motion of cracks appears unlikely. Only if very long range 
correlations in residual stresses 
existed would a quasi-static explanation be viable. However, 
even if these did occur, one would have replaced the problem of understanding 
the apparent universality of crack roughness exponents in a wide variety of 
materials with the problem of why residual stress correlations should be 
long range and universal.
A more appealing alternative is that elastodynamics of the medium plays 
an essential role. It has been shown that, in the case of a crack front 
restricted to a plane, the sound waves emitted as it moves changes its 
behaviour
 both when it is moving at a finite velocity and near the 
threshold\cite{unpublished}.
Such effects may also play a crucial role in increasing the roughness of 
the crack
surface\cite{langer}.

We finally note that in thin plates, our quasi-static analysis yields a 
crack path with a roughness exponent of $\zeta=1/2$ for 
tensile cracks and $\zeta=1$ for tearing cracks; in these situations 
elastodynamic effects may be less important, but buckling may play a role.

We would like to thank J.~R.~Rice, E.~Bouchaud and J.-P.~Bouchaud for 
useful discussions. This work has been supported in part by the NSF via 
DMR-9106237, 9630064 and Harvard University's MRSEC.

\end{document}